\documentclass[12pt]{article}
\usepackage{graphicx}
\usepackage{amssymb}
\usepackage{xcolor}

\definecolor{MYBlue}{rgb}{0.0470,0,0.5294}
\definecolor{MYRed}{RGB}{253,7,7}
\definecolor{MYGreen}{RGB}{10,250,10}

\begin{document}

\title{\bf Dynamical environments of relativistic binaries:
The phenomenon of resonance shifting}

\author{Ivan~I.~Shevchenko$^{1,2,3}$, Guillaume Rollin$^4$, Jos\'e Lages$^4$}

\date{}

\maketitle

\noindent $^1$Pulkovo Observatory, Russian Academy of Sciences,
196140 Saint Petersburg, Russia \\
email: {\tt iis@gaoran.ru}

\noindent $^2$Saint Petersburg State University, 7/9
Universitetskaya nab., 199034 Saint Petersburg, Russia

\noindent $^3$Lebedev Physical Institute, Russian Academy of
Sciences, 119991 Moscow, Russia

\noindent $^4$Institut UTINAM, Observatoire des Sciences de
l'Univers THETA, CNRS, \\ Universit\'e de Bourgogne
Franche-Comt\'e, Besan\c{c}on 25030, France

\date{}

\maketitle

\begin{center}
Abstract
\end{center}

\noindent In this article, we explore both numerically and
analytically how the dynamical environments of mildly relativistic
binaries evolve with increasing the general relativity factor
$\gamma$ (the normalized inverse of the binary size measured in
the units of the gravitational radius corresponding to the total
mass of the system). Analytically, we reveal a phenomenon of the
relativistic shifting of mean-motion resonances: on increasing
$\gamma$, the resonances between the test particle and the central
binary shift, due to the relativistic variation of the mean
motions of the primary and secondary binaries and the relativistic
advance of the tertiary's pericenter. To exhibit the circumbinary
dynamics globally, we numerically integrate equations of the
circumbinary motion of a test particle, and construct relevant
scans of the maximum Lyapunov exponents and stability diagrams in
the ``pericentric distance -- eccentricity'' plane of initial
conditions. In these scans and diagrams, regular and chaotic
domains are identified straightforwardly. Our analytical and
numerical estimates of the shift size are in a good agreement.
Prospects for identification of the revealed effect in
astronomical observations are discussed.

\bigskip


\section{Introduction}

Any gravitating binary with the mass parameter (ratio of masses of
the companions) greater than a specific threshold has a zone of
chaos around it, where all circumbinary orbits of low-mass
particles, irrespective of their eccentricities, are chaotic, due
to accumulation of integer mean-motion resonances to the parabolic
separatrix \cite{S15}. This underlines the importance of
consideration of circumbinary resonant phenomena in actual
astrophysical situations.

If any gravitating binary, whose dynamical environments are under
study, is massive and close enough, one has to take into account
effects of general relativity. Some aspects of resonant and
chaotic orbital dynamics around relativistic binaries were
explored by means of numerical experiments in
\cite{SM10,Seto13,HW14GRG}. In \cite{SM10}, an evolution of a
massive black hole binary with a resonantly trapped circumbinary
neutron star was studied. In \cite{Seto13}, a specific resonance
between the relativistic precession of a binary and the orbital
motion of a distant third body around the binary was analyzed.
Evolving chaotic dynamics of a test particle orbiting around a
decaying (due to the gravitational radiation) relativistic binary
was considered in \cite{HW14GRG}.

In this article, we focus on resonant circumbinary phenomena.
Namely, we explore the dynamical environments of mildly
relativistic binaries, with an emphasis on analyzing the resonant
structure of the border of the circumbinary chaotic zone.

A paradigmatic example of a relativistic binary is the black hole
binary GW150914 shrinking before the merger: a transient
gravitational-wave signal was observed by the Laser Interferometer
Gravitational-Wave Observatory (LIGO) in 2015; it was attributed
to the merger of two black holes \cite{A16a,A16b,A16c}. In the
source frame, the initial masses of the two merging black holes
were estimated to have been $m_1 \sim 36$ and $m_2 \sim 29$ Solar
masses \cite{A16a}. New discoveries in this rapidly developing
field of astronomy emerge \cite{A17PRL}; therefore, theoretical
studies of the pre-merger evolution, starting from mildly
relativistic phases, of such and similar objects are actual.
GW150914 represents just an example where circumbinary resonant
phenomena can be present. Although circumbinary matter was not
observed directly in the case of GW150914, it is inherent (in the
form of planets and disks) to close stellar binaries, including
binaries that have compact objects as components. Much more
details and examples are given below in Section~\ref{sec_Obs}.

In this article, we restrict our analysis to solely mildly
relativistic binaries. Analytically, we reveal a phenomenon of
resonance shifting in the dynamical vicinity of a relativistic
binary: on increasing $\gamma$ (the normalized inverse of the
binary size measured in the units of the gravitational radius
corresponding to the total mass of the system), the mean-motion
resonances between the test particle and the central binary shift
slowly, due to the relativistic variation of the mean motions and
the relativistic periastron advance. We derive formulas predicting
this phenomenon analytically, construct relevant scans of Lyapunov
exponents and stability diagrams, and compare the obtained
numerical results with our theoretical predictions.

The paper is organized as follows. In Sections~\ref{sec_lir} and
\ref{sec_rscr}, a general theoretical approach to the effect of
the relativistic shifting of circumbinary mean-motion resonances
is given. In Section~\ref{sec_em}, we describe our numerical
experiments, present Lyapunov exponent scans and stability
diagrams, and compare the numerical results with the theoretical
predictions. In Section~\ref{sec_Obs}, prospects for astronomical
observations of resonance shifts in the dynamics around
relativistic binaries are considered. Section~\ref{sec_Concl} is
devoted to the general discussion and conclusions.

\section{Resonant circumbinary dynamics}
\label{sec_lir}

In the given problem, mean-motion resonances \cite{MD99,Morbi02}
correspond to commensurabilities between the orbital frequencies
of the primary binary and an orbiting test particle. In the case
of the particle's circumbinary motion, one has $a > a_\mathrm{b}$,
where $a$ and $a_\mathrm{b}$ are the particle's and the binary's
semimajor axes, respectively. We consider mean-motion resonances
of the form $(k+q)/k$, where $k$ and $q$ are integers, and we set
$1 - k \leq q \leq -1$ and $k \geq 3$. Therefore, in what follows,
$q$ is always negative. The integer $| q |$ is the order of the
resonance.\footnote{Note that the letter $q$ is also used,
traditionally, to designate the pericentric distance.} Therefore,
the ratio of the particle' and binary's orbital frequencies is
close to $(k+q)/k = (k - |q|)/k$ and is less than one.

Extending the analysis presented in \cite{HM96,MH97}, the
Hamiltonian of a particle's motion in the vicinity of a
mean-motion particle-binary resonance $(k+q)/k$ in the restricted
elliptic planar three-body problem can be written as

\vspace{-3mm}

\begin{equation}
H = {1\over2} \beta \Lambda^2 - \sum_{p=q}^{p=0}\phi_{k+q, k+p,
k}\cos(\psi + p \varpi), \label{HM_cb}
\end{equation}

\noindent where $\beta={3k^2} / a^2$, $\Lambda = \Psi -
\Psi_\mathrm{res}$, $\Psi = (\mu_1 a)^{1/2}/k$, $\Psi_\mathrm{res}
= (\mu_1^2/(k^2(k+q)n_\mathrm{b}))^{1/3}$, $\mu_1 = 1 - \mu$, $\mu
= m_2/(m_1 + m_2)$ (we set $m_1 > m_2$); $\varpi$ is the longitude
of the particle's pericenter; $\psi = k l - (q+k) l_\mathrm{b}$,
where $l$ and $l_\mathrm{b}$ are the mean longitudes of the
particle and the binary. Note that index $p \leq 0$.

The units are chosen in such a way that the total mass of the
primary (central) binary, the gravitational constant, and the
primary binary's semimajor axis $a_\mathrm{b}$ are all equal to
one. The primary binary's mean longitude $l_\mathrm{b} =
n_\mathrm{b} t$, and the primary binary's mean motion
$n_\mathrm{b} = 1$, i.\,e., the time unit equals ${1 \over 2
\pi}$th part of the binary's orbital period.

Model~(\ref{HM_cb}) represents a truncated expansion of the
original Hamiltonian, as expressed in resonance canonical Delaunay
variables \cite{Morbi02}, in the Laplace series in the vicinity of
a given high-order ($|q| \gtrsim 2$) mean-motion resonance; the
expansion is truncated by ignoring the rapidly oscillating and
small-amplitude terms. In \cite{HM96,MH97}, the model was
demonstrated to provide a good description of the
close-to-resonant dynamical behavior, if its applicability
conditions are satisfied; see also comments to
formula~(\ref{cb_coeff}) below.

If the central binary is eccentric (the primary binary's
eccentricity $e_\mathrm{b} > 0$), resonance $(k+q)/k = (k -
|q|)/k$ splits in a cluster of $|q|+1$ subresonances $p = 0, -1,
-2, \ldots, q$. The resonant argument of each subresonance is
given by

\begin{equation}
\phi = \psi + p \varpi = k l - (q+k) l_\mathrm{b} + p \varpi = k l
- (k-|q|) l_\mathrm{b} - |p| \varpi ,
\label{res_argcb}
\end{equation}

\noindent where $p = 0, -1, -2, \ldots, q$.

At $q = 1 - k$, one has integer resonances (the binary's and
particle's periods are in the ratio $1/k$, while their mean
motions are in the integer ratio $k/1$). At $k \geq 2$, each of
them splits (if $e_\mathrm{b} > 0$) into a cluster of $k$
subresonances with the arguments

\begin{equation}
\phi = k l - l_\mathrm{b} + p \varpi
\label{res_argcbi}
\end{equation}

\noindent with $p = 0, -1, -2, \ldots, 1-k$.

The coefficients of the subresonant terms are given by

\vspace{-3mm}

\begin{equation}
|\phi_{k+q, k+p, k}| \approx {\mu \over { | q | \pi a}} {| q |
\choose | p |} \left(\epsilon \over 2 \right)^{| p |}
\left(\epsilon_\mathrm{b} \over 2 \right)^{| q | - | p |},
\label{cb_coeff}
\end{equation}

\noindent where $\epsilon = {{e a} / {|a-a_\mathrm{b}|}}$,
$\epsilon_\mathrm{b} = {{e_\mathrm{b} a} / {|a-a_\mathrm{b}|}}$;
$a_\mathrm{b}$ and $e_\mathrm{b}$ are the primary binary's
semimajor axis and eccentricity, respectively; $a$ and $e$ are the
particle's semimajor axis and eccentricity, respectively.

In any application, formula~(\ref{cb_coeff}) can be considered to
provide a satisfactory qualitative precision, if $\epsilon |q| <
1$ \cite{HM96}. Besides, as already noted above,
model~(\ref{HM_cb}) is restricted to the resonances of relatively
high order: $|q| \gtrsim 2$.

The frequency of small-amplitude oscillations on subresonance
$|p|$ is given by

\vspace{-3mm}

\begin{equation}
\omega_0 = (\beta|\phi_{k+q, k+p,k}|)^{1/2} \approx
{a\over{|a-a_\mathrm{b}|}} n_\mathrm{b} \left[ \mu_1 \mu {{4 |q|}
\over {3\pi}} {|q| \choose |p|} \left(a \over
{a_\mathrm{b}}\right) \left(\epsilon \over 2 \right)^{| p |}
\left(\epsilon_\mathrm{b} \over 2 \right)^{|q| - |p|}
\right]^{1/2} .
\label{cb_om0}
\end{equation}

According to Eq.~(\ref{cb_coeff}), if the central binary is
circular (its eccentricity is zero), only one subresonance
persists, the last one (that with $| q | = | p |$).

As it is clear, e.g., from Eq.~(\ref{res_argcbi}), the apsidal
precession may influence the location of resonances. Therefore,
let us consider the relevant effects invoking the precession in
the given dynamical configuration. There exists a number of causes
for the apsidal precession, among them general relativity. A
relatively rapid apsidal precession in our Solar system is
exhibited by Mercury. The rate of Mercury's apsidal precession due
to perturbations from all other planets is equal to
$532^{\prime\prime}$ per century; general relativity adds
$43^{\prime\prime}$ per century (see, e.g., \cite{Cl47}), whereas
the Solar oblateness and tidal effects are negligible.

Einstein's formula for the relativistic apsidal precession rate of
a particle orbiting around single central mass $M$ is

\begin{equation}
\omega_\mathrm{E} \equiv \dot \varpi = \frac{6 \pi \mathcal{G}
M}{c^2 a (1 - e^2)} = \frac{3 \pi R_{\mathrm{g}}}{q (1 + e)}
\label{pr_rel}
\end{equation}

\noindent (in radians per particle's orbital revolution)
\cite{Cl47}, where $\mathcal{G}$ is the gravitational constant,
$c$ is the speed of light, $a$ and $e$  are the semimajor axis and
eccentricity of the particle's orbit; the gravitational radius
$R_\mathrm{g}$ of the central mass $M$ is equal to $2 \mathcal{G}
M / c^2$.

For the classical non-relativistic case, approximate analytical
expressions describing the secular dynamics in the hierarchical
circumstellar and circumbinary systems can be found in
\cite{H78,MN04,DS15}. In the hierarchical circumbinary version of
the circular ($e_\mathrm{b} = 0$) restricted planar three-body
problem, the apsidal precession rate of a circumbinary passively
gravitating tertiary, in ratio to the particle's mean motion $n$,
is given by

\begin{equation}
\frac{\omega_\mathrm{cl}}{n} = \frac{3}{4} \frac{m_1 m_2}{(m_1 +
m_2)^{2}} \left( \frac{a_\mathrm{b}}{a} \right)^{2} = \frac{3}{4}
\mu (1-\mu) \left( \frac{a_\mathrm{b}}{q} \right)^{2} (1-e)^{2} .
\label{prec_clnb}
\end{equation}

\noindent Here the subindex ``cl'' of $\omega$ means
``classical'', i.e., the problem is non-relativistic. The
barycentric frame is adopted; $m_1$ and $m_2$ are the masses of
the binary components (we set $m_1 \ge m_2$); $\mu = m_2/(m_1 +
m_2)$; $a_\mathrm{b}$ and $e_\mathrm{b}$ are the primary binary's
semimajor axis and eccentricity; $a$, $q$, and $e$ are the
tertiary's semimajor axis, pericentric distance and eccentricity.

In circumbinary configurations, general relativity can contribute
much to the apsidal precession of the tertiary, if the central
binary is massive and close enough. In the hierarchical
circumbinary problem one has $M = m_1 + m_2$ in
Eq.~(\ref{pr_rel}); therefore

\begin{equation}
\frac{\omega_\mathrm{E}}{n} = 3 \gamma \frac{a_{\mathrm{b}}}{q (1
+ e)} , \label{pr_relcb}
\end{equation}

\noindent where the dimensionless parameter $\gamma$, as defined
in \cite{MD94}, is given by

\begin{equation}
\gamma = \frac{\mathcal{G} M_\mathrm{S}}{c^2 a_\mathrm{E}} \cdot
\frac{(m_1 + m_2)}{a_\mathrm{b}} = 9.870994 \cdot 10^{-9} \cdot
\frac{(m_1 + m_2)}{a_\mathrm{b}}, \label{eq_gamma}
\end{equation}

\noindent and $M_\mathrm{S}$ is the Solar mass, $a_\mathrm{E}$ is
the astronomical unit, $a_\mathrm{b}$ is the binary's size, $m_1$
and $m_2$ are in Solar units, $a_\mathrm{b}$ is in astronomical
units. From Eq.~(\ref{eq_gamma}) we see that the $\gamma$ factor
is just the normalized inverse of the size of the binary measured
in the units of the gravitational radius corresponding to the
total mass of the system.

In the post-Newtonian formalism, Eq.~(\ref{pr_relcb}) for the
pericenter advance is valid in its first approximation. For the
second PN approximation, relevant expressions can be found, e.g.,
in \cite{KG05PRD,KG06PRD,B13MN,MM08,BHS17,DVG18}. We adopt the
formula as given in \cite{KG06PRD} and in \cite{DVG18}, Eq.~(A1),
rewriting it in the form

\begin{equation}
\frac{\omega_\mathrm{1PN+2PN}}{n} = \frac{\omega_\mathrm{E}}{n} +
\left( \frac{\omega_\mathrm{E}}{n} \right)^2 \left( \frac{13}{6} +
\frac{17}{12} \, e^2 \right) ,
\label{pr_2PN}
\end{equation}

\noindent where $\omega_\mathrm{E}$ is given by
Eq.~(\ref{pr_relcb}). As one may conclude from this formula, the
2PN term in Eq.~(\ref{pr_2PN}) may start to compete with the 1PN
term at large $e$ and $\gamma$. Therefore, henceforth we restrict
our analysis to relatively small eccentricities $e \lesssim 0.2$.

The ratio of the relativistic (1PN) precession rate and the
classical precession rate, as follows from Eqs.~(\ref{prec_clnb})
and (\ref{pr_relcb}), is given by

\begin{equation}
\frac{\omega_\mathrm{E}}{\omega_\mathrm{cl}} = \frac{4 \gamma}{\mu
(1-\mu)} \frac{q}{a_\mathrm{b}} (1+e)^{-1} (1-e)^{-2} .
\label{pr_ratio}
\end{equation}

The relativistic apsidal precession around a gravitating binary
also contains a contribution due to the rotation of the binary;
this contribution is retrograde. If $m_2 \ll m_1$, the rate of the
precession induced by the spin angular momentum $L$ of the binary
is given by

\begin{equation}
\frac{\omega_L}{n} = - \frac{4 \mathcal{G} m_2
a_\mathrm{b}^{1/2}}{c^2 a^{3/2} (1 - e^2)^{3/2}} ,
\label{pr_relam}
\end{equation}

\noindent see Eqs.~(A2)--(A3) in \cite{YA12}, also \cite{LL62}.
Introducing $\gamma$ and $q$ in this formula, one has

\begin{equation}
\frac{\omega_L}{n} \simeq - 4 \mu \gamma \left(
\frac{a_\mathrm{b}}{q} \right)^{3/2} (1 + e)^{-3/2} .
\label{pr_relamcb}
\end{equation}

\noindent As revealed in \cite{YA12}, the rotating central binary
induces also a quadruple moment and the corresponding (much
smaller) prograde precession rate is given by

\begin{equation}
\frac{\omega_Q}{n} \simeq \left( 6 + \frac{171}{8} e^2 \right) \mu
\frac{a_\mathrm{b}^2}{a^3} = \left( 6 + \frac{171}{8} e^2 \right)
\mu \gamma \left(\frac{a_\mathrm{b}}{q} \right)^3 (1 - e)^3 ,
\label{pr_rel_YA12}
\end{equation}

\noindent with accuracy $\sim e^4$. Thus, the total rate of the
relativistic precession caused by the central binary is given by

\begin{equation}
\frac{\omega_\mathrm{rel}}{n} = \frac{\omega_\mathrm{E}}{n} +
\frac{\omega_L}{n} + \frac{\omega_Q}{n} . \label{pr_rel_sum}
\end{equation}

\section{Relativistic shifting of circumbinary resonances: theory}
\label{sec_rscr}

The relativistic shifts of the mean motion $n_1$ of the central
binary and the mean motion $n_2$ of the tertiary from their
nominal values $n_1^{(0)}$ and $n_2^{(0)}$ can be calculated using
Eqs.~(6)--(7) in \cite{MD94}; we rewrite these equations in the
form

\begin{equation}
n_1 = \left( 1 + \frac{\mu(1-\mu) - 3}{2} \gamma_1 \right)
n_1^{(0)} , \label{n1}
\end{equation}

\begin{equation}
n_2 = \left( 1 - \frac{3}{2} \gamma_2 \right) n_2^{(0)} ,
\label{n2}
\end{equation}

\noindent where $\gamma_1$ and $\gamma_2$ are the values of the
$\gamma$ parameter calculated via Eq.~(\ref{eq_gamma}), where
$a_\mathrm{b}$ is set equal, respectively, to the semimajor axis
of the central binary and to the semimajor axis of the tertiary's
orbit. Designating the ratio of the orbital periods of the
tertiary and the primary binary by $f_T$, one has for the value of
this ratio shifted from the nominal location $f_T^{(0)}$:

\begin{equation}
f_T \approx \left[ 1 + \frac{\mu(1-\mu)}{2} \gamma + \frac{3}{2}
\left( \frac{1-e}{q} - 1 \right) \gamma \right] f_T^{(0)} ,
\label{fT}
\end{equation}

\noindent where $\gamma \equiv \gamma_1$ and $q$ is measured in
units of the primary binary size. Here it was taken into account
that the correction is much less than the ratio itself.
Designating the ratio of the semimajor axes by $f_a$, one has

\begin{equation}
f_a \approx \left[ 1 + \frac{\mu(1-\mu)}{3} \gamma + \left(
\frac{1-e}{q} - 1 \right) \gamma \right] f_a^{(0)} . \label{fa}
\end{equation}

\noindent For $e \sim 0$ and $q \gg 1$, these two formulas reduce
to

\begin{equation}
f_T \approx \left( 1 - \frac{3}{2}\gamma \right) f_T^{(0)}
\label{fTe0}
\end{equation}

\noindent and

\begin{equation}
f_a \approx \left( 1 - \gamma \right) f_a^{(0)} . \label{fae0}
\end{equation}

Let us now calculate how the shift of resonance can be influenced
by the relativistic apsidal precession. We consider the integer
mean-motion resonances (the binary's and particle's periods are in
the ratio $1/k$, while their mean motions are in the integer ratio
$k/1$).

As follows from Eq.~(\ref{cb_coeff}), in the circular
($e_\mathrm{b}=0$) restricted planar three-body problem only one
subresonance is present in the mean-motion resonance multiplet,
namely, that with $p=1-k$. The resonance is therefore not split
and is represented in the sum~(\ref{HM_cb}) solely by the term
with $p=1-k$. For the system to be in resonance, the averaged time
derivative of the resonant argument, given by
Eq.~(\ref{res_argcbi}), should be equal to zero: $k \dot{l} -
\dot{l_\mathrm{b}} + (1-k) \dot{\varpi} = 0$. Therefore, the
location of an integer mean-motion resonance is given by

\begin{equation}
\frac{n_\mathrm{res}}{n_\mathrm{b}} = \frac{1}{k} + \frac{k-1}{k}
\cdot \frac{\omega_\Sigma}{n_\mathrm{b}}, \label{eq_nres}
\end{equation}

\noindent where the apsidal precession rate is the sum of the
classical and relativistic contributions:

\begin{equation}
\omega_\Sigma = \omega_\mathrm{cl} + \omega_\mathrm{rel} .
\label{eq_omsigma}
\end{equation}

In the following, we are interested in the resonance relativistic
shift from the classical nominal location affected already by the
classical circumbinary precession. Based on
Eq.~(\ref{eq_omsigma}), it is straightforward to assess the amount
of this shift:

\begin{equation}
f_T \approx \left[ 1 - (k-1) \frac{\omega_\mathrm{E}}{n} \right]
f_T^{(0)} \approx \left[ 1 - \frac{3 (k-1) \gamma }{q (1 + e)}
\right] f_T^{(0)} , \label{fTrel}
\end{equation}

\noindent where $q$ is measured in units of the primary binary
size. Analogously,

\begin{equation}
f_a \approx \left[ 1 - \frac{2 (k-1) \gamma }{q (1 + e)} \right]
f_T^{(0)} . \label{farel}
\end{equation}

\noindent At $e \sim 0$, one has

\begin{equation}
f_T \approx \left[ 1 - \frac{3 (k-1) \gamma }{q} \right] f_T^{(0)}
\label{fTrele0}
\end{equation}

\noindent and

\begin{equation}
f_a \approx \left[ 1 - \frac{2 (k-1) \gamma }{q} \right] f_a^{(0)}
. \label{farele0}
\end{equation}

\noindent The total shift in the semimajor axis is given by the
sum of the shifts in (\ref{fa}) and (\ref{farel}):

\begin{equation}
\Delta f_a \equiv \frac{f_a}{f_a^{(0)}} - 1\approx
\frac{\mu(1-\mu)}{3} \gamma + \left( \frac{1-e}{q} - 1 \right)
\gamma - \frac{2 (k-1) \gamma }{q (1 + e)} ,
\label{sumfarel}
\end{equation}

\noindent or, at $e \sim 0$ and $q \gg 1$,

\begin{equation}
\Delta f_a \approx - \left[ 1+ \frac{2 (k-1)}{q} \right] \gamma .
\label{sumfarele0}
\end{equation}

\noindent For example, assume that $e=0$, $q = 2$, and $k=3$, then
from Eq.~(\ref{sumfarel}) follows $\Delta f_a \approx - 2.5
\gamma$ and from Eq.~(\ref{sumfarele0}) follows $\Delta f_a
\approx - 3 \gamma$.

Note that, as already mentioned above, in the case when the
primary binary is circular ($e_\mathrm{b} = 0$), solely the
$p=1-k$ term is present in the multiplet; others are non-existent
(have zero coefficients) and, therefore, there is no need to
consider them. Concerning half-integer and other higher-level (in
the Farey tree defined below in Section~\ref{sec_em}) resonances,
their shifts can be estimated as averages taken over their two
integer neighbors in the Farey tree.

Concluding this Section, one may summarize that the mean-motion
resonances between the tertiary and the primary binary are subject
to shifting from their nominal locations due to (1)~the
relativistic corrections to the mean motions of the central binary
and the tertiary, and (2)~the relativistic apsidal precession of
the tertiary's orbit. In both cases, the shift in the period
ratio or in the semimajor axis ratio is negative and is of the
order of $\gamma$; thus, the cumulative effect is also of the
order of $\gamma$. The total shift can be calculated using
Eq.~(\ref{sumfarel}) or (\ref{sumfarele0}).

\section{Stability diagrams}
\label{sec_em}

Our numerical simulations are performed in the framework of the
circular restricted three-body planar problem in post-Newtonian
approximation, adopting the equations of motion as given in
\cite{MD94}. The software code developed in \cite{MD94} is used.

To distinguish between regular and chaotic types of motion in the
given problem, we use a statistical method proposed and applied in
\cite{MS98,SM03}. It consists of four steps. ({\it i}) On a
representative set of initial data, two differential distributions
(histograms) of the orbits in the computed value of $\log_{10}
\lambda_\mathrm{max}$ (where $\lambda_\mathrm{max}$ is the maximum
Lyapunov exponent) are constructed, using two different time
intervals for the integration. ({\it ii}) If the phase space of
motion is divided \cite{C79}, each of these distributions has at
least two peaks. The peak that shifts (moves in the negative
direction in the horizontal axis), when the integration time
interval is increased, corresponds to the regular orbits. The peak
(or a set of peaks) that stays still corresponds to the chaotic
orbits. ({\it iii}) The value of $\log_{10} \lambda_\mathrm{max}$
at the histogram minimum between the peaks is identified, thus
providing a numerical criterion for separating the regular motion
from the chaotic one. ({\it iv}) The obtained criterion can be
used in any further computations to identify regular/chaotic
orbits on much finer initial data grids and, rather often, on
smaller time intervals of integration. For the case of
circumbinary motion, this method was implemented in \cite{PS13} to
construct stability diagrams of circumbinary planetary systems in
the ``$q$--$e$'' (pericentric distance --- eccentricity) space of
initial conditions of planetary orbits.

Henceforth, we adopt the following designations: $T_\mathrm{r}$ is
the time of simulation (in orbital periods of the central binary),
$q$ is the initial pericentric distance (in units of the central
binary size), $\mu \equiv m_2/(m_1 + m_2)$ is the mass parameter
of the central binary ($m_2 < m_1$), $\gamma$ is the relativistic
factor already defined.

\begin{figure}[h!]
\centering
\includegraphics[width=0.8\textwidth]{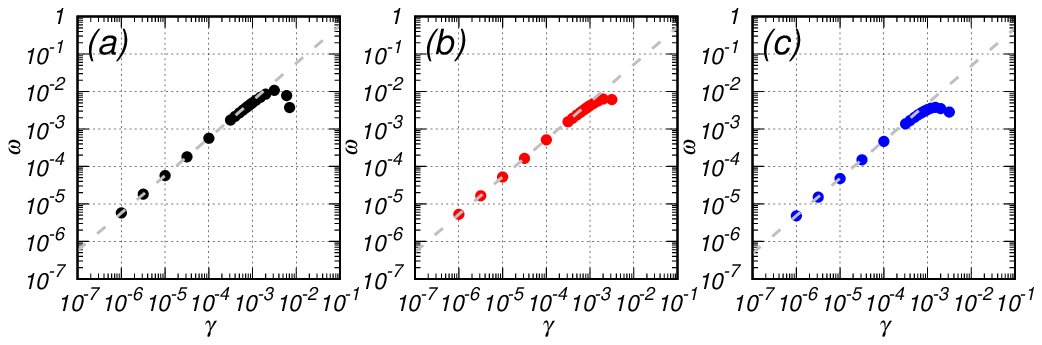}
\caption{The rate of apsidal precession of the particle's orbit
around a single primary, as a function of the relativistic factor
$\gamma$. For all panels, we set the particle's pericentric
distance $q = 3$. The eccentricities: (a) $e = 0.1$, (b) $e =
0.2$, (c) $e = 0.3$. The dots represent the results of our
numerical simulations, and the dashed straight line is the
theoretical $\omega = 2 \pi \frac{\omega_\mathrm{E}}{n}$, where
$\frac{\omega_\mathrm{E}}{n}$ is given by Eq.~(\ref{pr_relcb}).}
\label{Fig_adv}
\end{figure}

In Fig.~\ref{Fig_adv}, we illustrate the limits of validity of the
code used. The rate of precession of the pericenter of a
particle's orbit around a single primary is shown as a function of
the relativistic factor $\gamma$. (For a single primary, the
relativistic factor $\gamma$ is formally defined by setting the
mass of the secondary to zero and the binary size to unity.) For
all panels of the Figure, the particle's pericentric distance $q =
3$. The eccentricities are: (a) $e = 0.1$, (b) $e = 0.2$, (c) $e =
0.3$. The dots represent the results of our numerical simulations,
and the dashed straight line is the theoretical $\omega = 2 \pi
\frac{\omega_\mathrm{E}}{n}$, where $\frac{\omega_\mathrm{E}}{n}$
is given by Eq.~(\ref{pr_relcb}).

One can see that the performance of the code improves with $e$
decreasing: while at $e=0.3$ the code sharply starts to produce a
false retrograde precession at $\gamma > 0.003$, at $e=0.1$ the
precession starts to be retrograde much later, at $\gamma >
0.008$. Therefore, one may estimate and expect that at $e \approx
0$ the code is valid at $\gamma$ values up to $0.01$.

\begin{figure}[h!]
\centering
\includegraphics[width=0.8\textwidth]{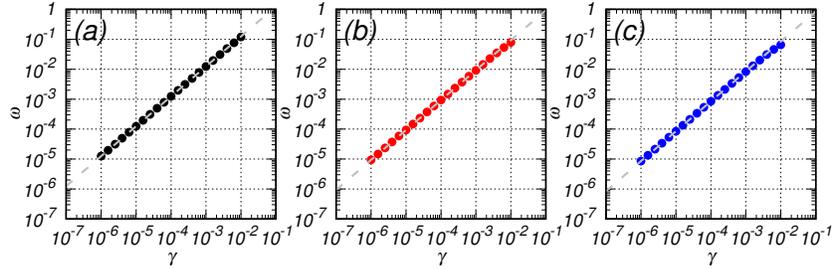}
\caption{The same as in Fig.~\ref{Fig_adv}, but at $e=0.01$ (all
panels) and $q=1.5$ (panel (a)), $q=2.0$ (panel (b)), $q=2.2$
(panel (c)).}
\label{Fig_adv_esmall}
\end{figure}

Our computations, performed separately at three relevant points of
the diagram, namely at ($q =1.5$, $e=0.01$), ($q =2.0$, $e=0.01$),
and ($q =2.2$, $e=0.01$), confirm this expectation. In
Fig.~\ref{Fig_adv_esmall}, the precession rate dependence on
$\gamma$ is shown for these three cases. The curves are apparently
in a good agreement with the 1PN theory up to $\gamma$ values as
large as 0.01.

\begin{figure}[h!]
\centering
\includegraphics[width=0.8\textwidth]{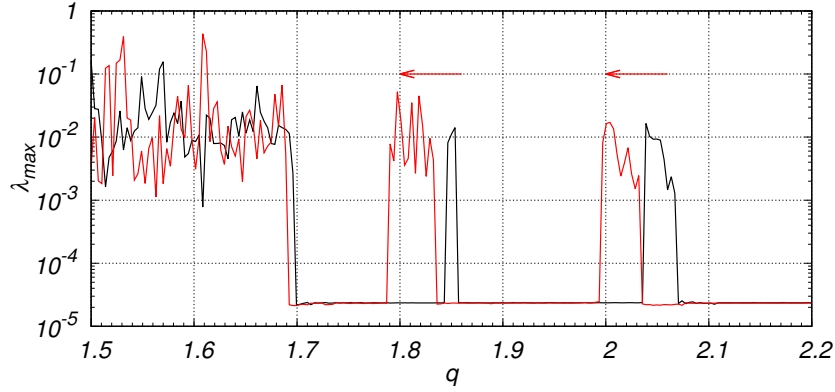}
\caption{Scans of the maximum Lyapunov exponent along the $q$ axis
at $e=0.01$. The black curve corresponds to $\gamma=10^{-7}$, and
the red one to $\gamma=10^{-2}$. The red arrows indicate the
resonance shift direction.}
\label{Fig_scan}
\end{figure}

Now let us proceed to a numerical verification of the resonance
shift phenomenon in the three-body problem. We set $\mu=0.1$ and
build two scans (one for $\gamma \approx 0$ and one for $\gamma
=0.01$) of the maximum Lyapunov exponent along the $q$ axis at a
small fixed $e$, namely, at $e=0.01$. The scan's interval covers a
relevant neighborhood of $q \sim 2$, namely, $q \in [1.5, 2.2]$.
Note that panels (a) and (c) of Fig.~\ref{Fig_adv_esmall}
correspond to the borders of this interval. The scans are
presented in Fig.~\ref{Fig_scan}. The black curve corresponds to
$\gamma=10^{-7}$, and the red one to $\gamma=10^{-2}$. The 3/1
resonance chaotic band (at $q \approx 2$) is apparently shifted by
the amount of $\approx -0.03$. As already calculated in
Section~\ref{sec_rscr} by means of Eq.~(\ref{sumfarel}), the
theoretical $\Delta f_a \approx - 2.5 \gamma \approx - 0.03$. The
perfect agreement of the numerical-experimental result with the
theory is evident; and the resonant shift phenomenon is thus
confirmed numerically.

\begin{figure}[h!]
\centering
\includegraphics[width=0.8\textwidth]{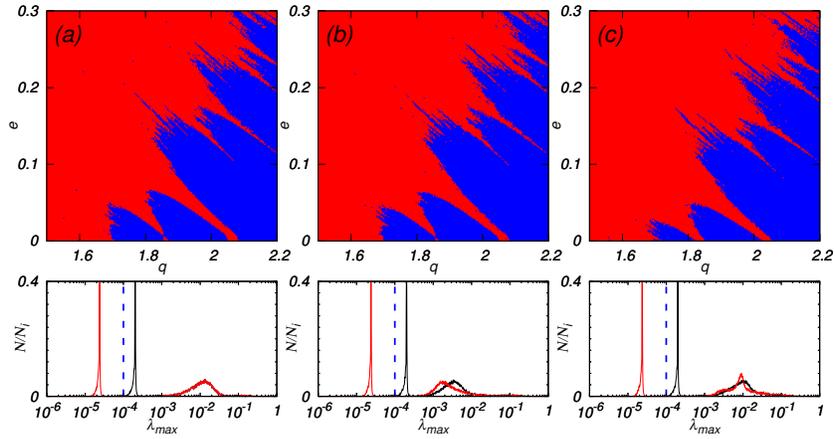}
\caption{Upper panels: ``$q$--$e$'' stability diagrams at fixed
$\mu=0.1$. Chaotic and regular zones are shown in red and blue,
respectively. Panel~(a): $\gamma=10^{-50} \sim 0$; panel~(b):
$\gamma=10^{-3}$; panel~(c): $\gamma=10^{-2}$. Panels below: the
corresponding histograms of computed values of Lyapunov exponents;
see the text for details. } \label{Fig_stab}
\end{figure}

To provide a global picture of the circumbinary dynamics, we
construct ``$q$--$e$'' stability diagrams (Fig.~\ref{Fig_stab}) in
the given problem. In the upper panels of Fig.~\ref{Fig_stab}, the
stability diagrams are presented for various $\gamma$ values at
fixed $\mu=0.1$. Chaotic and regular zones in the plane of initial
values of $e$ and $q$ are shown in red and blue, respectively.

These diagrams are presented here just for illustrative purposes,
to provide a global overview of resonances. The used code is not
accurate enough to characterize resonance shifts at $\gamma =
0.01$ and $e = 0.1$--0.3 (i.e., in the upper part of the third
panel in the Figure). What is more, the deviations between the
three panels cannot be used to quantify our effect, because it was
described above analytically for the case of $e \approx 0$.
However, at $e \approx 0$, the code is expected to perform
accurately enough.

To distinguish between regular and chaotic orbits, the statistical
method \cite{MS98,SM03} is used, as described above. Namely, two
values of the finite-time maximum Lyapunov exponent
$\lambda_\mathrm{max}$ are computed on a grid of initial values on
two time intervals $T_\mathrm{r}=10^4$ and $T_\mathrm{r}=10^5$
(measured in the central binary revolutions); then, the
$\lambda_\mathrm{max}$ distributions are compared. The panels
below the stability diagrams in Fig.~\ref{Fig_stab} demonstrate
the corresponding histograms of the maximum Lyapunov exponent
$\lambda_\mathrm{max}$. The $N/N_i$ value is the normalized
relative number of orbits with the given $\lambda_\mathrm{max}$.
The histograms are computed on fine grids of initial $q$ and $e$,
as prescribed above by the general algorithm of constructing such
diagrams. Black and red curves correspond to the two computation
times of $10^4$ and $10^5$ binary periods, respectively. The
derived numerical criteria for the separation between regular and
chaotic orbits are indicated in the histograms by dashed vertical
lines. It is apparent that the fixed and shifted peaks are
well-separated (no overlap of them is present) and thus the
regions of chaos and order are well-defined.

The chaos border in panels (a), (b), and (c) is ragged due to
resonances; the most prominent ``teeth'' (bands) correspond to
integer resonances (the binary's and particle's periods are in the
ratio $1/k$, while their mean motions are in the integer ratio
$k/1$). The Farey tree \cite{Meiss92} of the resonant teeth at the
border is evident. Recall how the Farey tree is built: consider
first the lowest order ``neighboring'' ratios $m/n$ and $m'/n'$
(in the given case, these are the integer ratios $m/1$ and
$(m+1)/1$); then, the next (higher) level of the Farey tree is
made of mediants given by the formula $m''/n''= (m + m')/(n + n')
= (2m+1)/2$. Thus the half-integer mean-motion resonances are the
mediants for the integer ones, and so on. The orbital resonances
accumulate more and more densely with increasing $k$, i.e., on
approaching the parabolic separatrix; this effect is evident in
the upper right parts of the panels in Fig.~\ref{Fig_stab}. Note
that the shape of the resonant bands is essentially sensitive to
variations of the system parameters; this is generic for marginal
resonances; see \cite{S12}. Thus, the diagrams in
Fig.~\ref{Fig_stab} graphically demonstrate how major resonances
interact and overlap.

As already mentioned above, at $e \approx 0$ the code is expected
to perform accurately enough, and the resonance shifting
phenomenon can be checked. Consider the resonant ``tooth,''
directed to the point ($q \approx 2$, $e=0$). It corresponds to
the mean-motion resonance 3/1. Comparing diagrams~(b) and (c) (at
$e \approx 0$, where the code is accurate), one may see that, on
increasing $\gamma$ by amount of 0.01, the tooth is shifted at
this point to the left in the semimajor axis, by amount of
$\approx - 0.03$. Therefore, in this diagram, the numerical result
again agrees with the analytical estimate already given above.

\begin{figure}[h!]
\centering
\includegraphics[width=0.5\textwidth]{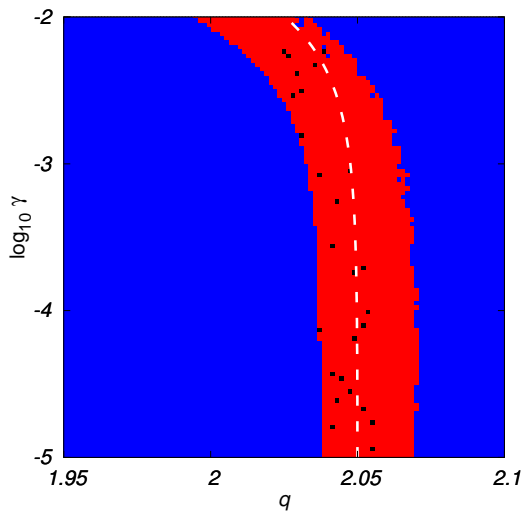}
\caption{The stability diagram $\gamma$ vs $q$ in the neighborhood
of the 3/1 resonance. Chaotic and regular zones are shown in red
and blue, respectively. The theoretical curve is white dashed.}
\label{Fig_gvq}
\end{figure}

\begin{figure}[h!]
\centering
\includegraphics[width=0.8\textwidth]{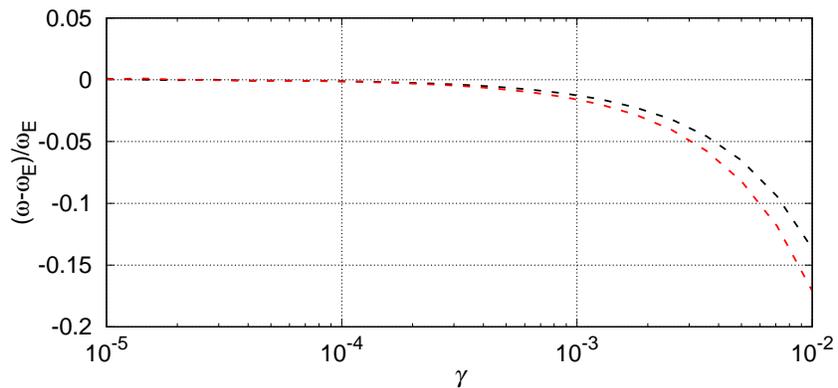}
\caption{The normalized difference $(\omega - \omega_\mathrm{E}) /
\omega_\mathrm{E}$ as a function of $\gamma$. The numerical code
is used to compute $\omega$, and $\omega_\mathrm{E}$ is calculated
by Eq.~(\ref{pr_relcb}). The black and red curves correspond to
$q=1.95$ and $q=2.05$, respectively; $e=0.01$.}
\label{Fig_diff_om}
\end{figure}

To illustrate the effect in a more general setting, in
Fig.~\ref{Fig_gvq} we present a stability diagram constructed in
the plane of values of the $\gamma$ parameter and the initial
value of $q$; whereas $\mu=0.1$, $e=0.01$. The method of
construction is the same as for Fig.~\ref{Fig_stab}; the technical
histogram of regular and chaotic orbits, not shown here, again
demonstrates their perfect separation. In Fig.~\ref{Fig_gvq}, the
vicinity of the 3/1 resonance is displayed. Chaotic and regular
zones are shown in red and blue, respectively. The white dashed
curve is given by Eq.~(\ref{sumfarel}) at $k=3$, $\mu=0.1$,
$e=0.01$, $q=2.051$. We see that the resonance shift increases
with $\gamma$ and it generally follows the theoretical relation.
At the highest values of $\gamma$ ($\sim 0.01$) in the diagram,
the computed shift seems to be somewhat stronger than the
theoretical one. This may be due to restrictions in both the
theory and simulation code. To reveal graphically the code
validity limits, in Fig.~\ref{Fig_diff_om} we present the
normalized difference $(\omega - \omega_\mathrm{E}) /
\omega_\mathrm{E}$ as a function of $\gamma$. Here $\omega$ is
computed using the numerical code and $\omega_\mathrm{E}$ is
calculated by Eq.~(\ref{pr_relcb}). The presented curves
correspond to $q=1.95$ and $q=2.05$ (in the vicinity of the 3/1
resonance, see Fig.~\ref{Fig_gvq}); $e=0.01$. An artifact
deviation of up to $\approx 16$\% is seen at $\gamma = 0.01$. This
deviation in $\omega$ may result in a deviation of up to $\approx
12$\% in the total resonance shift at $\gamma = 0.01$, but the
theory limitations come already into play at this level, as, e.g.,
we do not take into account the role of the binary's angular
momentum. Finally, we may conclude that the theory and simulation
results are in a reasonable quantitative agreement.

\section{Observability of the effect}
\label{sec_Obs}

What are the prospects for astronomical observations of such
resonance shifts in the dynamics of matter around relativistic
binaries? Generally, circumbinary material is now routinely
observed in the form of planets and disks \cite{W14IAU,P18HP}.
Near-resonant circumbinary planets (CBPs) were identified
\cite{D11S,PS13}, as well as resonant features in circumbinary
disks; see, e.g., \cite{DS16MN}. On the other hand, planetary
systems of compact stars are known to be an ordinary phenomenon
\cite{P18HP}. Moreover, historically, the first exoplanet was
discovered in 1992 in a system of a neutron star (hereafter NS)
\cite{WF92N}. Nowadays, the ubiquity of planets in orbits around
such compact objects as single pulsars is well established. Tight
binaries with compact components such as white dwarfs (hereafter
WD) are observed that have CBPs \cite{P18HP}; a particular example
is NN~Ser.

As soon as the considered effect is proportional to $\gamma$, let
us see, first of all, which binaries may have values of $\gamma$
large enough. As given by Eq.~(\ref{eq_gamma}), the $\gamma$
factor is just the normalized inverse of the size of the binary
measured in the units of the gravitational radius corresponding to
the total mass of the system. Therefore, high values of $\gamma$
are inherent to binaries that are tight and massive.

Merging black hole binaries, already mentioned in the
Introduction, such as GW150914, have $\gamma \sim 1$ at their late
pre-merger stages of evolution; but any low-mass material around
them (such as planets) is practically impossible to observe at the
present technological level. There may exist relevant
observationally verifiable mechanisms, such as production of
free-floating planets, but considering such possibilities is far
beyond the scope of our article.

Among the ordinary stellar binaries, the tightest ones are the
so-called contact binaries, the W~UMa-type stars. For them, the
binary mass is $\sim 2$ Solar masses, and the binary size is $\sim
2$ Solar radii; therefore $\gamma \sim 10^{-6}$. This value should
be typical for yellow and red dwarf contact binaries, because for
small main-sequence stars (M~dwarfs) mass is approximately
directly proportional to radius; see Table~1 in \cite{KT09}.

If we consider WD--WD and NS--NS binaries in close-to-contact
configurations (not yet observed), we find that $\gamma$ would be
equal to $\sim 10^{-4}$ for a WD contact binary and $\sim 0.1$ for
an NS contact binary (given that WD and NS masses are of order of
a Solar mass and their radii are $\sim 10^4$ and $\sim 10$~km,
respectively). These are the values that can be expected at late
stages the inspiralling evolution of such binaries. If any
close-to-resonant circumbinary planet were observed in such a
system at this stage, its resonant shift can be identified by
modern methods of exoplanetary studies, such as ETV (eclipse
timing variation) or TTV (transit timing variation) methods in the
case of WD--WD systems, and the pulsar timing method in the case
of NS--NS systems. (For descriptions of the methods see
\cite{P18HP}.) If any disk material were present, shifts of
resonant features can also be searched for.

In principle, even for ordinary contact binaries (W~UMa-type
stars) the effect can be searched in the near future, because the
TTV method already provides the relative precision of $\sim
10^{-4}$ in the determination of orbital periods of CBPs; see
examples in \cite{D11S,W12N}.

Apart from stellar binaries, another relevant class of
astrophysical objects, perspective from the observational
viewpoint, is represented by supermassive black hole (SMBH)
binaries in active galactic nuclei, such as OJ~287 and
PKS~1302-102. The SMBH masses in active galactic nuclei can be as
great as $\sim 10^{10}$ in Solar units, whereas the sizes of the
SMBH binaries in OJ~287 and PKS~1302-102 are $\sim 0.1$~pc in the
both cases \cite{SHV88,GDS15}. Therefore, $\gamma \sim 0.01$ for
both OJ~287 and PKS~1302-102. Resonance shifts can be searched in
the dynamics of any material orbiting around the SMBH binaries, if
such material were identified.

Concluding, two major classes of astrophysical objects, namely,
(1)~contact or close-to-contact stellar binaries and (2)~SMBH
binaries in active galactic nuclei, represent perspective targets
for astronomical observation and verification of the considered
effect.

\section{Discussion and conclusions}
\label{sec_Concl}

In this article, we have explored both numerically and
analytically how the dynamical environments of mildly relativistic
binaries may evolve with increasing the general relativity factor
$\gamma$.

To exhibit the circumbinary dynamics globally, we have
constructed, using direct numerical integrations, the
``$q$--$e$''stability diagrams of the circumbinary orbits, in
which any resonant and chaotic features can be straightforwardly
identified.

Both analytically and numerically, we have revealed a new
phenomenon of the relativistic shifting of mean-motion resonances:
on increasing $\gamma$, the main chaotic resonant bands
(corresponding to integer resonances between the test particle and
the central binary) shift, due to the relativistic corrections to
the mean motions of the primary and secondary binaries and due to
the relativistic advance of the tertiary's pericenter. We have
derived formulas describing this phenomenon analytically. Our
analytical and numerical estimates of the shift size agree well.

Apart from the theoretical and numerical-experimental aspects of
the resonance shift phenomenon, we have considered prospects for
its astronomical observation and verification. We find that two
major classes of astrophysical objects, namely, (1)~contact or
close-to-contact stellar binaries and (2)~SMBH binaries in active
galactic nuclei, are perspective for identifying the effect in
observations.

Finally, we note that any numerical-experimental exploration of
the resonant dynamical environments on further increasing $\gamma$
requires much harder numerical simulation efforts, exploiting
codes taking into account higher PN approximations. We leave this
exploration for a future work. However, extrapolating the
resonance shifting phenomenon to the domain of larger values of
$\gamma$, i.e., physically, to harder relativistic inspiralling
binaries, one may expect that, on increasing $\gamma$, the
resonant bands would further shift slowly closer to the central
binary. Graphically, this phenomenon might be called a ``sundew''
effect, in analogy to a carnivorous plant that looks similar to
the presented stability diagrams. The ``sundew'' effect might be
important for the fate of any circumbinary matter (planets,
planetesimals, dust, or dark matter particles), if present. The
dynamical regular evolution of the matter between the slowly
closing resonant teeth, as well as the chaotic evolution inside
them, deserves a further numerical study.

\section*{Acknowledgements}

We express our deep gratitude to the referee, whose remarks helped
to improve the manuscript. We are most grateful to Thomas Maindl
for supplying us with the relativistic three-body integrator code.
It is a pleasure to thank Dima Shepelyansky for helpful comments.
I.I.S. was supported in part by the Russian Foundation for Basic
Research (project No.~17-02-00028) and by the Programmes of
Fundamental Research of the Russian Academy of Sciences
``Nonlinear dynamics: Fundamental problems and applications'' and
`Problems of origin and evolution of the Universe with application
of methods of ground-based observations and space research''
(CP19--270).

\newpage

\vspace{15mm}

\newpage

\vspace{15mm}

\newpage

\vspace{15mm}

\end{document}